\documentclass[%
 prd,
superscriptaddress,twocolumn,
 amsmath,amssymb,
]{revtex4-2}

\usepackage[dvips]{graphicx}
\usepackage{epsfig}
\usepackage{hyperref}
\usepackage{amsmath}
\usepackage{url}
\usepackage{color}
\usepackage{soul}

\begin{document}

\title{Planck Formula for the Gluon Parton Distribution in the Proton}

\author{Loredana Bellantuono}
\email{loredana.bellantuono@ba.infn.it}
\affiliation{Dipartimento di Scienze Mediche di Base, Neuroscienze e 
Organi di Senso, Universit\`a degli Studi di Bari Aldo Moro, 
I-70124, Bari, Italy}
\affiliation{Istituto Nazionale di Fisica Nucleare, Sezione di Bari, 
I-70125, Bari, Italy}

\author{Roberto Bellotti}
\affiliation{Dipartimento Interateneo di Fisica, Universit\`a degli Studi di 
Bari Aldo Moro, I-70126, Bari, Italy}
\affiliation{Istituto Nazionale di Fisica Nucleare, Sezione di Bari, 
I-70125, Bari, Italy}

\author{Franco Buccella}
\affiliation{INFN, Sezione di Napoli, Via Cintia, Napoli, I-80126, Italy} 


\begin{abstract}
We describe the gluon parton distribution function (PDF) in 
the proton, deduced by data from the ATLAS and HERA experiments, in the 
framework of the parton statistical model. The best fit parameters involved 
in the Planck formula that describes the gluon distribution are consistent 
with the results obtained from analysis of deep inelastic scattering processes.
Remarkably, the agreement between the statistical model and the experimental 
gluon distributions is found with the same value of the ``temperature'' 
parameter $\bar{x}$ found by fitting the valence parton distributions 
from deep inelastic scattering. This result corroborates the validity of the 
statistical approach in the gluon sector.
\end{abstract}

\maketitle

\section{Introduction}

The scale invariance in deep inelastic scattering of leptons (electrons, muons
and neutrinos) on nucleons \cite{BP} had a crucial role in the proposal of 
quantum chromodynamics (QCD) \cite{FGL} as the field theory of strong 
interactions. To describe the phenomenon, Feynman \cite{F} proposed that the 
hadrons behave at large $Q^2$ as an incoherent set of point-like objects, 
called \textit{partons}, characterized by a given probability of carrying a 
fraction $x\in[0,1]$ of the hadron momentum in the rest frame of the final 
hadrons. While the charged partons have been identified with the quarks 
\cite{LS}, a relevant fraction of the hadron momentum is carried by neutral 
partons, identified with the gluons, that play the role of gauge bosons of 
QCD. The $Q^2$ dependance of the parton distributions is implied by QCD, as 
described by the DGLAP equations \cite{GL,D,AP}, which have been experimentally confirmed \cite{A}. 
Therefore, if one fixes an initial $Q^2_0$, sufficiently high that 
perturbative QCD is reliable for larger values, the parton distributions 
can be derived as a function of $Q^2$ by the DGLAP equations. 

Polynomial functions are often considered for the boundary conditions, 
but theoretical ideas inspired by experimental facts suggest a different 
parametrization. The idea that Pauli principle implies $\bar{d}(x)$ larger 
than $\bar{u}(x)$ \cite{NS,FF} has been confirmed by the defect \cite{NMC} 
in the Gottfried sum rule \cite{G} and by the experiments on Drell-Yan 
production of pairs in proton-proton and proton-deuteron scattering 
\cite{NA,Do}. This fact has inspired to write parton distributions for the 
boundary low-$Q^2$ conditions \cite{BST} of DGLAP equations according to 
quantum statistical mechanics \cite{BBS} in the variable $x$, which appears 
in the parton model sum rules. The ``potentials'' that appear in the Fermi-Dirac
distributions of the valence partons depend on flavor ($q=u,d$) and helicity 
($h=+,-$). This feature determines the intriguing possibility to describe both
the unpolarized fermion distributions $q(x) = q^{+}(x) + q^{-}(x)$ and their 
polarized counterparts $\Delta q(x) =q^{+}(x) - q^{-}(x)$ \cite{BBS,BBS2,BBS3}. 

An important constraint to the model is the hypothesis that, at the separation
 between the non-perturbative and the perturbative QCD regimes, there is 
equilibrium \cite{Bh1,Bh2,Bh3} for the elementary processes involved in the 
DGLAP equations. As a consequence, gluons must be described by a Planck 
formula, namely a Bose-Einstein distribution with vanishing chemical potential, 
while the isospin and spin asymmetries of the sea are related to the 
non-diffractive contribution to the valence parton distributions. 
This feature allows to predict, in agreement with the experiment (see Refs.~\cite{Do,Ada}),
\begin{equation}\label{sea}
\Delta \bar{d}(x) < 0 < \Delta \bar{u}(x) < \bar{d}(x) - \bar{u}(x) <
\Delta \bar{u}(x) - \Delta \bar{d}(x) .
\end{equation}
Deep inelastic processes are unable to probe the gluon distribution with 
precision, since gluons, that are singlets with respect to the electroweak 
group, appear in the logarithmic correction of the parton distributions of the fermions. 
Instead, they play an important role, as an octet of $SU(3)_c$, in the 
strong $p-p$ interactions measured at ATLAS. Purpose of this article is 
to describe the parton distribution deduced by the measurements at ATLAS 
\cite{ATLAS} and HERA \cite{HERA} in the light of the statistical model. More specifically, we will determine the parameters of 
the statistical gluon distribution by fitting experimental data, and 
compare the result with the outcomes of previous studies, which were 
obtained from constraints on QCD sum rules.

The article is organized as follows: in Section~\ref{sec:model}, we introduce the statistical model and briefly discuss previous findings concerning both the fermion and gluon sectors; in Section~\ref{sec:gluons}, we perform the fit of the gluon distribution function; in Section~\ref{sec:conclusions}, we summarize the results and discuss their relevance.

\section{Parton statistical model}\label{sec:model}

The parameters of the statistical model found in the seminal work \cite{BBS} 
at $Q^2=4\,\mathrm{GeV}^2/c^4$ were also successfully used to describe the 
polarized nucleon structure functions \cite{BBS2,BBS3}. Following studies 
determined the same parameters at $Q^2 = 1\,\mathrm{GeV}^2/c^4$ \cite{BoS} 
and by comparison \cite{BSo} with the parton distributions proposed in 
Ref.~\cite{HERA}. We report in Table \ref{tab:tab1} the 
values, found in the aforementioned studies, of the relevant parameters 
characterizing the quark and antiquark distributions \cite{BBS}
\begin{align}
x q^h(x) & = \frac{A X_{q}^h x^b}{\exp[(x-X_{q}^h)/\bar{x}]+1} 
+\frac{\tilde{A} x^{\tilde{b}}}{\exp(x/\bar{x}) + 1} , \\
x \bar{q}^h(x) & = \frac{\bar{A} (X_{q}^{-h})^{-1} x^{2b}}
{\exp[(x+X_{q}^{-h})/\bar{x}]+1} 
+\frac{\tilde{A} x^{\tilde{b}}}{\exp(x/\bar{x}) + 1} ,
\end{align}
(with $q=u,d$ and $h=+,-$) and the gluon distribution \cite{BBS}
\begin{equation}\label{gluons}
x g(x) = \frac{A_G x^{b_G}}{\exp(x/\bar{x}) - 1} .
\end{equation}
The factors in the first terms of the fermion distributions may be
explained by the extension to the transverse degrees of freedom
\cite{BBS4} exactly for $X_{q}^h$ and approximately for 
$(X_{q}^{-h})^{-1}$. The comparison among the parameters found in Refs.~\cite{BBS,BoS,BSo} shows 
stability for $\bar{x}$, which we denote as the ``temperature'' of the model, 
and for the ``potentials'' $X_{q}^h$ of the valence partons, depending on 
their flavor and helicity. Instead, the parameters $A_G$ and $b_G$, which 
appear in the Planck distribution \eqref{gluons} of the gluons, are 
characterized by a more striking variability. The same occurs for the 
parameters $\tilde{A}$ and $\tilde{b}$ that determine the diffractive 
term of the fermion distributions. The discrepancy of Ref.~\cite{BoS} with 
respect to Ref.~\cite{BBS} is due to the choice of a smaller $Q^2$, 
lying in a region where the gluon and diffractive distributions are expected 
to become narrower as a consequence of scale dependence. In the case of 
Ref.~\cite{BSo}, differences are due to the fact that the parameters of the 
statistical model were fixed to match the distributions proposed in 
Ref.~\cite{HERA}. 
In fact, the factor $(1 - x)^C$ of the standard 
parametrization and the Boltzmann factor $\exp(-x/\bar{x})$ 
have a different behaviour, as stressed in Ref.~\cite{BST}, where the 
statistical description has been shown to be in a better agreement with the 
gluon distribution found in Ref.~\cite{NNPDF}. 

It is thus crucial to determine the gluon distribution
measured at ATLAS \cite{ATLAS}. In fact, while in deep 
inelastic scattering with incident leptons, gluons, that 
are singlets with respect to the electroweak gauge group, 
are fixed by their role in the DGLAP equations \cite{D,GL,AP}, 
in proton-proton scattering they interact strongly as color octets. 
Therefore, one can hope to gain more information on the gluon 
distribution from LHC experiments. For this reason, we compare the 
prediction of the statistical approach with the experimental 
values, where they do not depend on the extrapolation following from 
the parametrization.

\begin{table}
\centering
\begin{tabular}{l|rrr}
Parameter & ~\cite{BBS} & ~\cite{BoS} &  ~\cite{BSo}  \\  
\hline
$\bar{x}$   &   0.099  &   0.090  &  0.099 \\
$X_{u}^+$ & 0.461 & 0.475 & 0.446 \\ 
$X_{u}^-$ & 0.298 & 0.307 & 0.297 \\
$X_{d}^+$ & 0.228 & 0.245 & 0.222 \\
$X_{d}^-$ & 0.302 & 0.309 & 0.320 \\   
$A_G$ & 14.3 & 32.8 & 27.18 \\
$b_G$ & 0.747 & 1.02  & 0.75 \\
$\tilde{A}$ & 1.91  & 0.147 & 0.07 \\
$\tilde{b}$ & $- 0.253$ & 0.043 & $-0.25$ \\
\hline
\end{tabular}
\caption{Values of the statistical model parameters found in previous works.
The temperature $\bar{x}$ is involved in both the fermion and gluon 
distributions. The ``potentials'' $X_{u}^+$, $X_{u}^-$, $X_{d}^+$ and 
$X_{d}^-$ determine the non-diffractive parts of the fermion distributions, 
while $\tilde{A}$ and $\tilde{b}$ fix the diffractive ones. 
Finally, $A_G$ and $b_G$ appear in the gluon distribution.}
\label{tab:tab1}
\end{table}

The difference for the gluon and the diffractive terms may be the 
consequence of a different value of $Q^2$ chosen in Ref.~\cite{BoS},
since the distributios are modified by the evolution, and by the
comparison with the very different parametrization of HERA for
the gluons: in fact the comparison with NNPDF \cite{NNPDF} shows
a better agreement for Eq.~\eqref{gluons}. 

\section{Fit of the gluon distribution function}\label{sec:gluons}

\begin{figure}
    \centering
    \includegraphics[width=0.48\textwidth]{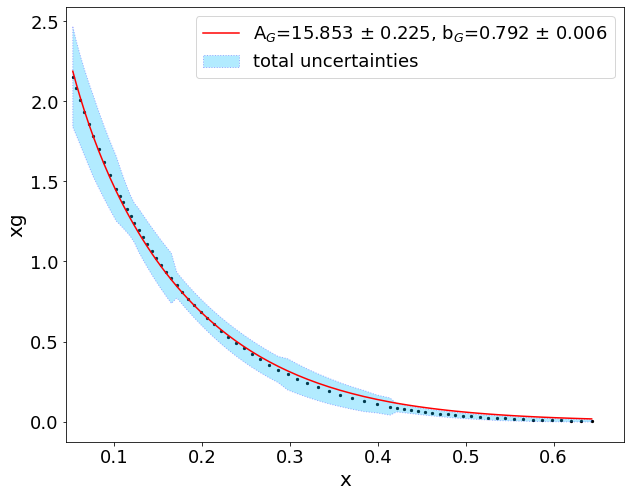}
    \caption{The red curve represents the best fit of the 
gluon momentum distribution $xg(x)$ obtained in the ATLAS experiment, 
performed using the functional form in Eq.~\eqref{gluons}, with $A_G$ and 
$b_G$ as free parameters, and $\bar{x}=0.099$. The dots correspond to the 
experimental points, and the (cyan) shaded area to their uncertainty.}
    \label{fig:gluons}
\end{figure}

Since we assume 
$Q^2_0 = 4\,\mathrm{GeV}^2/c^4$, we choose to limit 
the $x$ range to $[0.05,0.66]$, where the lower bound is fixed 
to avoid the QCD corrections proportional to 
$\alpha_s(Q^2_0) \left|\ln{x}\right|$, while the upper bound 
ensures that the invariant mass:
\begin{equation}\label{invariantmass}
(M')^2 = M^2_p + Q_0^2 \frac{1-x}{x} ,
\end{equation}
with $M_p$ the proton mass, is not too small. We choose to fit the central 
values of the gluon momentum distribution $xg(x)$ obtained in the ATLAS 
experiment with free parameters $A_G$ and $b_G$, while we fix 
$\bar{x}=0.099$, since it was determined in Ref.~\cite{BBS} by the rapidity 
of the decrease of the fermionic  parton distributions around their 
``potentials'' and above them. The best fit of $N=74$ points in 
$[0.05,0.66]$ provides
\begin{equation}\label{gluon_parameters}
A_G = 15.853 \pm 0.225, \quad b_G = 0.792 \pm 0.006 . 
\end{equation}
The agreement of this result with the experimental points is reported in 
Fig.~\ref{fig:gluons}. The error bars therein are determined by three 
kinds of uncertainties, that are summed in quadrature: the first kind is obtained from variations of the experimental parameters; the second is related to uncertainties in the physical constants, such as the quark mass and the coupling constants, of the physical model; the third one is determined by variations in the form of the fit function that provides the central value of the gluon distribution function in the ALTAS experiment \cite{ATLAS}. 
The visual agreement between the best fit and the reference points is 
confirmed by the value of $\chi^2=49.03$, which, considering the number 
$\nu=N-2=72$, corresponds to a significance level close to $0.975$.  

The agreement with the values 
\begin{equation}\label{gluon_parameters_old}
A_G = 14.28, \quad b_G = 0.75,
\end{equation}
found in Ref.~\cite{BBS}, is satisfactory, considering that in the 
aforementioned work the parameters of the gluon distribution were 
not fixed directly by a best fit operation, but rather on theoretical 
grounds, with $b_G$ determined 
by the parameter $\tilde{b}$, appearing in the diffractive light-quark 
contribution, as $b_G = \tilde{b} + 1$, and $A_G$ obtained from the momentum 
sum rule. The fact that similar parameters to the ones in 
Eq.~\eqref{gluon_parameters_old} are obtained by fitting the ATLAS data for 
the gluon distribution is remarkable, and confirms the validity of the 
statistical approach.

\begin{table}
    \centering
    \begin{tabular}{c|c|c|c|c}
         \ & This work & Ref.~\cite{BBS} & HERA~\cite{HERA} & Ref.~\cite{BST} 
\\ \hline
         $M_{g,\mathrm{low}}$ & $0.358 \pm 0.012$ & 0.37 & 0.34 & 0.32 \\
         $M_{g,\mathrm{high}}$ & $0.084 \pm 0.002$ & 0.08 & 0.05 & 0.13 \\
         $M_{g}$ & $0.442 \pm 0.014$ & 0.45 & 0.39 & 0.45 \\
         \hline
    \end{tabular}
    \caption{Integrated gluon momentum distributions for low ($x<0.2$) and 
high ($x>0.2$) values, for the model obtained in our work and for previous 
works.}
    \label{tab:momentum_integrals}
\end{table}

We report in Table~\ref{tab:momentum_integrals} the values of the integrated 
momentum distribution for low ($x<0.2$) and high ($x>0.2$) momentum values,
\begin{equation}
    M_{g,\mathrm{low}} = \int_0^{0.2} x g(x) dx, 
\quad M_{g,\mathrm{low}} = \int_{0.2}^1 x g(x) dx,
\end{equation}
along with the total momentum 
\begin{equation}
    M_{g} = \int_0^1 x g(x) dx = M_{g,\mathrm{low}} + M_{g,\mathrm{high}}
\end{equation}
carried by the gluons, using the best fit values \eqref{gluon_parameters} 
and $\bar{x}=0.099$. The comparison with the values obtained in 
Refs.~\cite{BBS,HERA,BST} confirms that the prediction of the gluon 
distribution fit is in a very good agreement with the one obtained in 
Ref.~\cite{BBS} and the shape significantly differs from the one 
proposed by Ref.~\cite{HERA}. 
More specifically there is good agreement at small $x$, but the difference 
becomes striking with increasing $x$. This effect is a consequence of the 
different behavior of the functions $\exp(-x/\bar{x})$ and $(1-x)^C$, 
that predict the large $x$ gluon distributions in the statistical 
and in the standard approach .
The good agreement with data of the Planck formula is a good point in 
favor of the statistical approach, considering furthermore that it is 
obtained considering the value of $\bar{x}$ found from the form of the 
Fermi-Dirac function for the non-diffractive term of the valence partons. 

\section{Conclusions}\label{sec:conclusions}

The good agreement of the Planck formula for the gluon parton distribution 
in the proton with experiment  with the same value for the ``temperature'', 
$\bar{x} = 0.099$, and the other parameters, $A_G$ and $b_G$ near to the 
ones found in the study of deep inelastic about twenty years ago, is a 
good point in favor of the parametrization inspired by quantum statistical 
mechanics, which allows a reliable extrapolation to the $x$ regions, where 
one has not a sufficient information from experiment.
The proposal of boundary conditions for the DGLAP equations 
Ref.~\cite{D,GL,AP} fixed by statistical quantum mechanics 
Ref.~\cite{BST} ispired by the role of Pauli principle advocated 
in Ref.~\cite{NS} and in Ref.~\cite{FF} receives an important confirmation 
from the study of ATLAS data Ref.~\cite{ATLAS}.

\begin{acknowledgments}
\textit{Acknowledgments.--} We are very grateful to Francesco 
Giuli for informing us about the results reported in Ref.~\cite{ATLAS} and 
for providing the gluon distributions at $Q^2 = 4 \mathrm{GeV}^2/c^4$. 
We thank Werner Vogelsang for the results reported in Ref.~\cite{Ada}.
\end{acknowledgments}

\end{document}